\documentclass[useAMS,usenatbib]{mn2e}
\usepackage{graphicx,amssym}
\newcommand{\bc}{\begin{center}}
\newcommand{\ec}{\end{center}}

\title[AGN in High Mass Galaxies]
      {Why are AGN found in High Mass Galaxies?}
\author[L.Wang, G.Kauffmann]
       {Lan Wang$^{1,}$$^2$$\thanks{Email: wanglan@mpa-garching.mpg.de}$,
        Guinevere Kauffmann$^2$
        \\
    $^1$Department of Astronomy, Peking University, Beijing 100871, China\\
        $^2$Max--Planck--Institut f\"ur Astrophysik,
        Karl--Schwarzschild--Str. 1, D-85748 Garching, Germany}
\begin{document}

\date{Accepted 2008 ???? ??.
      Received 2008 ???? ??;
      in original form 2008 ???? ??}

\pagerange{\pageref{firstpage}--\pageref{lastpage}} \pubyear{2008}

\maketitle

\label{firstpage}

\begin{abstract}
We use semi-analytic models implemented in the {\it Millennium
Simulation} to analyze the merging histories of dark matter haloes
and of the galaxies that reside in them. We assume that supermassive
black holes only exist in galaxies that have experienced at least
one major merger. Only a few percent of galaxies with stellar masses
less than $M_* < 10^{10} M_{\odot}$ are predicted to have
experienced a major merger and to contain a black hole. The fraction
of galaxies with black holes increases very steeply at larger
stellar masses. This agrees well with the observed strong mass
dependence of the fraction of nearby galaxies that contain either
low-luminosity (LINER-type) or higher-luminosity (Seyfert or
composite-type) AGN. We then investigate when the major mergers that
first create the black holes are predicted to occur. High mass
galaxies are predicted to have formed their black holes at very
early epochs. The majority of low mass galaxies never experience a
major merger and hence do not contain a black hole, but a
significant fraction of the supermassive black holes that do exist
in low mass galaxies are predicted to have formed recently.
\end{abstract}

\begin{keywords}
   galaxies: interactions -- galaxies: haloes -- galaxies: nuclei
\end{keywords}

\section{Introduction}
\label{sec:intro}

By studying active galactic nuclei (AGN), we learn about the physical
mechanisms that trigger accretion onto the central supermassive black
holes of galaxies. When a black hole accretes, it increases
in mass. By studying populations of AGN at low and at high redshifts,
we hope to infer the history of how black holes build up their mass.

It has been established that supermassive black holes most occur in
galaxies with bulges \citep{kormendy1995}, and that the mass of the
black hole correlates with the luminosity and the stellar velocity
dispersion of the host bulge \citep{magorrian1998, ferrarese2000,
gebhardt2000}. This indicates that the formation of galaxies and
supermassive black holes are likely to be closely linked. In the
local Universe, the fraction of bulge-dominated galaxies hosting AGN
decreases at lower stellar masses \citep{ho1997, kauffmann2003}. In
order to form a black hole, it is necessary for gas to lose angular
momentum and sink to the centre of the galaxy \citep{haehnelt1993,
volonteri2003}. The gravitational torques that operate during
galaxy-galaxy mergers are known to be a very effective mechanism for
concentrating gas at the centers of galaxies \citep{mihos1996}.
Models for AGN evolution have often assumed that black holes are
formed and fuelled, and AGN activity is triggered during major
mergers of galaxies \citep{kauffmann2000, wyithe2003, croton2006}.

At low and moderate redshifts, there is no conclusive observational
evidence that mergers play a significant role in triggering AGN
activity in galaxies. In the local Universe, \citet{li2006} have
shown that narrow line AGN do not have more close companions than
matched samples of inactive galaxies. Even at intermediate redshifts
(z $\sim 0.4-1.3$), moderate luminosity AGN hosts do not have
morphologies indicative of an ongoing merger or interaction
\citep{hasan2007}. The conclusion seems to be that although major
mergers may be responsible for AGN activity in some galaxies, other
fueling mechanisms are likely to be most important in the low
redshift Universe. It has also been established that high mass black
holes have largely stopped growing at early cosmic epochs, whereas
low mass black holes are still accreting at significant rates today
\citep{heckman2004}. X-ray observations show that very
high-luminosity AGN activity peaked at early cosmic epochs ($z \sim
2$), while low-luminosity AGN activity peaks at lower redshifts
\citep{steffen2003,barger2005,hasinger2005}.

It has been postulated that this so-called ``anti-hierarchical'' growth
of supermassive black holes can be explained if there are two modes of
accretion onto black holes that have very different efficiencies
\citep{merloni2004, mueller2007}. The early formation formation of ``new''
black holes may result in very luminous quasar-like events.
To form a supermassive black hole, a more violent process such as
a major merger may be required to funnel a large amount of gas into the
central region of the galaxy. Subsequent accretion of gas onto already
existing black holes may be an inefficient process and produce lower
luminosity AGN \citep{haehnelt1993, duschl2002}.
The history of accretion after the black hole is formed may not necessarily
be tightly linked to the dynamical history of the galaxy, but may be
controlled by the accretion and feedback processes occurring
in the vicinity of the black hole itself.

In this work, we use the combination of the {\it Millennium
Simulation} and semi-analytic models of galaxy formation to study
the fraction of galaxies that have undergone major mergers as a
function of mass and cosmic epoch. We investigate whether this can
be related to the demographics of black holes in the local Universe
and to the apparent disappearance of the most luminous quasar
activity in massive galaxies 
at late times.

In Sec.~\ref{sec:simulation}, we briefly introduce the simulation we use
and explain how galaxy mergers are tracked in the simulation.
In Sec.~\ref{sec:merger}, we show that if we assume that black holes only
form when galaxies undergo major merging events, then most present-day
low mass galaxies are predicted not not to contain black holes and hence
will not host AGN. In Sec.~\ref{sec:firstmerger}, we use the simulations
to predict when galaxies of different masses have underdone their first
major merger. Conclusions and discussions are presented in the final section.

\section{simulation and merger trees}
\label{sec:simulation}

The \emph{Millennium Simulation}\citep{springel2005} is used in this
work to study the merging histories of dark matter haloes. The
merging histories of galaxies can be inferred when the simulation is
combined with semi-analytic models that follow gas cooling, star
formation, supernova and AGN feedback and other physical processes
that regulate how the  baryons condense into galaxies.

The \emph{Millennium Simulation}  follows $N= 2160^3$ particles of mass
$8.6\times10^{8}\,h^{-1}{\rm M}_{\odot}$ from redshift $z=127$ to the
present day, within a comoving box of $500\, h^{-1}$Mpc on a side. The
cosmological parameters values in the simulation are consistent with the
determinations from a combined analysis of the 2dFGRS\citep{colless01} and
first year WMAP data \citep{spergel03}. A flat $\Lambda$CDM cosmology is
assumed with $\Omega_{\rm m}=0.25$, $\Omega_{\rm b}=0.045$, $h=0.73$,
$\Omega_\Lambda=0.75$, $n=1$, and $\sigma_8=0.9$.

Full particle data are stored at 64 output times. For each output,
haloes are identified using a friends-of-friends (FOF) group-finder.
Substructures (or subhaloes) within a FOF halo are located using the
SUBFIND algorithm of \citet{springel2001}. The self-bound part of
the FOF group itself also appear in the substructure list. This main
subhalo typically contains 90 percent of the mass of the FOF group.
After finding all substructures in all the output snapshots, subhalo
merging trees are built that describe in detail how these systems
merge and grow as the universe evolves. Since structures merge
hierarchically in CDM universes, for any given subhalo, there can be
several progenitors, but in general each subhalo only has one
descendant. Merger trees are thus constructed by defining a unique
descendant for each subhalo. We refer below {\em halo} to the main
substructure that can represent the FOF halo, while {\em subhalo}
refers to substructure other than the main one. Halo merger happens
when two FOF group merge into one group and one of the haloes
becomes a subhalo of the larger structure.

The substructure merger trees form the basic input to the
semi-analytic model used to associate galaxies with haloes/subhaloes
\citep{delucia2007}. The semi-analytic galaxy catalogue we are using
in this study is publicly available. A description of the publicly
available catalogues, and a link to the database can be found at the
webpage: http://www.mpa-garching.mpg.de/millennium/. Once a halo
appears in the simulation, a (central) galaxy begins to form within
it. The central galaxy is located at the position of the most bound
particle of the halo. As the simulation evolves, the halo may merge
with a larger structure and become a subhalo. The central galaxy
then becomes a satellite galaxy in the larger structure. The
galaxy's position and velocity are specified by the position and
velocity of the most bound particle of its host halo/subhalo. Even
if the subhalo hosting the galaxy is tidally disrupted, the position
and velocity of the galaxy is still traced through this most bound
particle. Galaxies thus only disappear from the simulation if they
merge with another galaxy. The time taken for a galaxy without
subhalo to merge with the central object is given by the time taken
for dynamical friction to erode its orbit, causing it to spiral into
the centre and merge. This is calculated using the standard
Chandrasekhar formula. All the information about the formation and
merging history of galaxies is stored.

By analyzing these halo and galaxy merger trees, we are able to
track when two haloes merge together and whether the galaxies within
them also merge into a single object by the present day. In this
study, we focus on mergers between satellite and central galaxies,
and exclude mergers between two satellites. These events are rare
\citep{springel2001} and neglecting them should not affect our
conclusions about the incidence and fueling of black holes in
galaxies.


\section{halo and galaxy mergers }
\label{sec:merger}

\begin{figure*}
\bc
\hspace{-1.6cm}
\resizebox{12.cm}{!}{\includegraphics{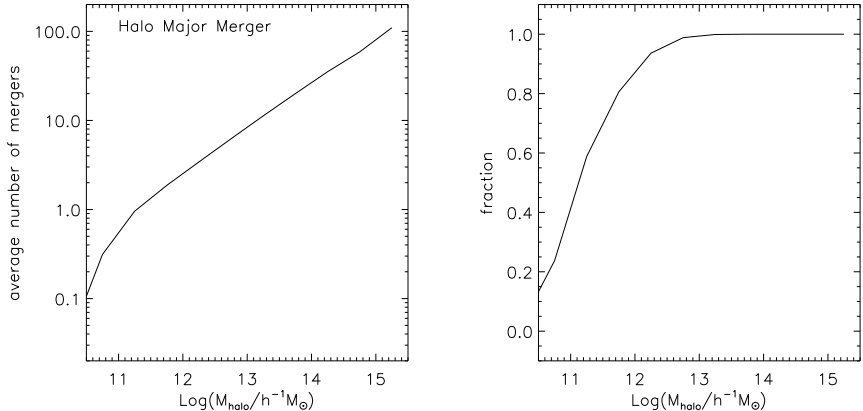}}\\%
\caption{
Left panel: the average number of major mergers that a dark matter halo
of given mass has experienced over its lifetime.
Right panel: the fraction of haloes of given
mass that have had at least one major merger.}
\label{fig:halofrac}
\ec
\end{figure*}

\begin{figure*}
\bc
\hspace{-1.6cm}
\resizebox{15.cm}{!}{\includegraphics{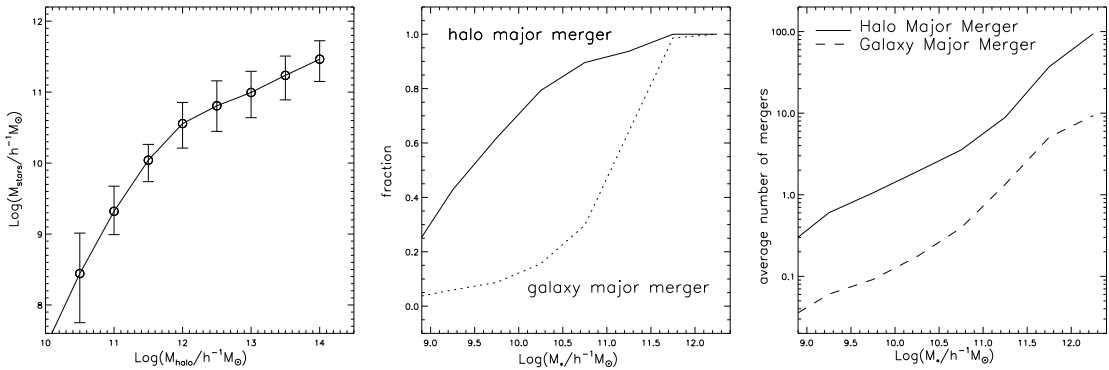}}\\%
\caption{ Left panel: The relation between the stellar mass of the
central galaxy and the the mass of its host dark matter halo as
predicted by the semi-analytic models of De Lucia \& Blaizot (2007).
The error bars indicate the $95$ percentile range in stellar mass at
a given value of $M_{halo}$ . Middle panel: The solid line shows the
fraction of dark matter haloes that have experienced at least one
major merger as a function of the stellar mass of the central
galaxy. The dotted line shows the fraction of central galaxies of
given mass that have had at least one major merger. Right panel: The
solid line shows the average number of major mergers experienced by
a dark matter halo as a function of the stellar mass of its central
galaxy. The dashed line shows the average number of major mergers
experienced by the central galaxy itself.} \label{fig:haloGal} \ec
\end{figure*}

\begin{figure}
\bc \hspace{-0.8cm}
\resizebox{9.cm}{!}{\includegraphics{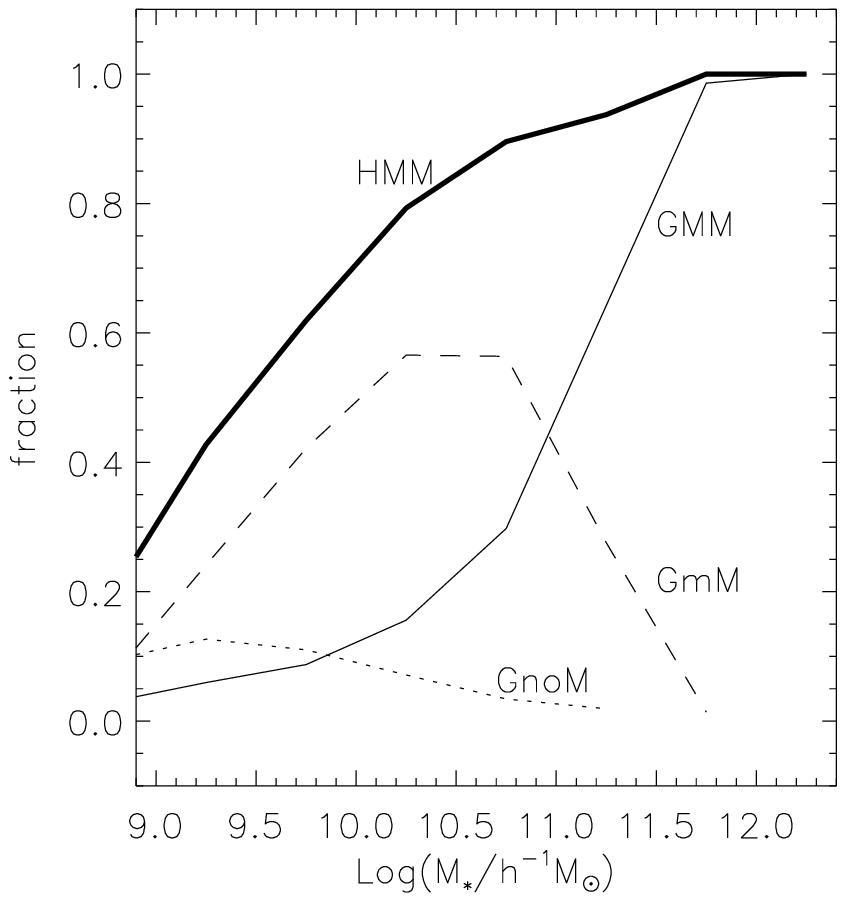}}\\%
\caption{ The thick solid line (HMM) shows the fraction of central
galaxies whose progenitor haloes have had at least one major merger.
The other lines split this sample of central galaxies according to
the history of the central galaxy itself. The dotted line (GnoM)
shows the contribution from central galaxies that have not
experienced a mergers of any kind. The dashed line (GmM) shows the
contribution from central galaxies that have experienced only minor
mergers. The thin solid line (GMM) shows the contribution from
central galaxies that have experienced major majors.}
\label{fig:Galfrac} \ec
\end{figure}

In this study, we assume that black holes form when a galaxy
undergoes a major merging event. Galaxies that have never
experienced a major merger do not have a black hole. We define major
mergers as events in which the mass ratio of the two progenitors is
greater than $0.3$. For halo merger, the mass ratio is the virial
mass ratio of two progenitor haloes. For galaxy merger, it is the
stellar mass ratio of two progenitor galaxies. When we track mergers
in the simulation, we include major mergers that occur in all
branches of the tree, not just the ``main branch''.

Since galaxies reside in dark matter haloes and are able to merge
only once their host haloes have coalesced, we begin by analyzing
the merging histories of the dark matter haloes themselves. In the
left panel of Fig.~\ref{fig:halofrac}, we plot the average number of
major mergers a present day dark matter halo has experienced over
its lifetime as a function of halo mass.  Note that in this analysis
we track mergers down to an effective resolution limit of $20$
particles, which corresponds to a halo of mass $1.7 \times
10^{10}h^{-1}M_{\odot}$.
We see that the number of major mergers (above the resolution limit)
experienced by a halo is a strongly increasing function of mass;
haloes with present-day masses of $10^{12} M_{\odot}$ have typically
experienced only one one major merger, whereas the progenitors of
present-day haloes with  masses of $10^{15} M_{\odot}$ have merged
with each other close to 100 times.

In the right panel, we show the fraction of haloes that have had at
least one major merger during their lifetime, as a function of halo
mass. The fraction of haloes that have had major mergers also
increases rapidly with halo mass. Almost all haloes more massive
than $10^{13}h^{-1}M_{\odot}$ have had at least one major merger and
this fraction drops to around 20 \% for haloes with masses of around
$10^{11}h^{-1}M_{\odot}$.

We now investigate the fraction of {\em galaxies} that have had
major mergers. The results are shown as a dotted line in the middle
panel of Fig.~\ref{fig:haloGal}. Rather than rising steeply as a
function of mass, the galaxy major merger fraction remains close to
zero up to a stellar mass of $10^{10.5} M_{\odot}$ and then rises
sharply. This is somewhat surprising in view of the behaviour of the
same quantity for dark matter haloes, plotted in the right-hand
panel of Fig.~\ref{fig:halofrac}. For reference, we have plotted the
relation between the stellar mass of a central galaxy and the mass
of its host halo in the left panel of Fig.~\ref{fig:haloGal}, as
predicted by the semi-analytic models we use in this study
\citep{delucia2007}. This mean relation can be used to transform
between central galaxy mass and halo mass in an approximate way
(this conversion neglects scatter between the two quantities and the
fact that some galaxies are actually satellite systems). If the
fraction of galaxies with major mergers followed the relation
derived for their host haloes, this would yield the solid curve in
the middle panel of Fig.~\ref{fig:haloGal}. Why are  the merging
histories of galaxies and their host haloes so different?

Once two dark matter haloes merge, the galaxies inside them will
merge together over a timescale that is determined by dynamical
friction. Upon investigation, we find that nearly all galaxies that
have experienced major mergers are located in dark matter haloes
that have also experienced a major merger. There are almost no
galaxy major mergers that have occurred in a halo that has only
experienced a minor merger ($\sim 0.15$ percent). However, the
converse is not true; {\em we find that a substantial fraction of
halo major mergers give rise to galaxy minor mergers.} This is
illustrated in the right-hand panel of Fig.~\ref{fig:haloGal}. The
solid line shows the number of major mergers experienced by the
progenitor {\em haloes} of a present-day central galaxy as a
function of its mass. The dashed line shows the number of major
mergers experienced by their progenitor {\em galaxies}. As can be
seen, the number of major mergers experienced by the progenitor
galaxies is an order of magnitude smaller. Notice that the number of
galaxy mergers is less than $1$ for galaxies up to $\sim
10^{11}h^{-1}M_{\odot}$, and increase steeply for massive galaxies.
This is in nice agreement with what is shown in Fig.9 of
\citet{lucia2006}, which shows the number of effective progenitors
as a function of the stellar mass for elliptical galaxies.

In Fig.~\ref{fig:Galfrac}, we again plot the fraction of central
galaxies of a given mass whose progenitor haloes have had a major
mergers (thick solid line). The thin solid line shows the fraction
whose progenitor {\em galaxies} have had a major merger. The dashed
line shows the fraction of such galaxies that have had minor mergers
and the dotted line is the fraction that have had no merger of any
kind. The main conclusion that can be gleaned from this plot is that
the reason why the thick solid and thin solid curves differ in
shape, is because at lower stellar masses, major mergers between the
progenitor haloes often correspond to minor mergers between the
progenitor galaxies.

How can we understand this? During the period of time between the
merger of the two haloes and the merger of the galaxies within them,
the stellar mass of the smaller ``satellite'' galaxy remains about
the same because ongoing star formation is quenched when the gas
surrounding the galaxy is shock--heated and no longer cools onto the
satellite. The central galaxy, however, will continue to increase in
mass as a result of cooling and star formation. The stellar mass
ratio of two galaxies therefore becomes smaller as a function of
time.

\begin{figure*}
\bc \hspace{-1.6cm}
\resizebox{15.cm}{!}{\includegraphics{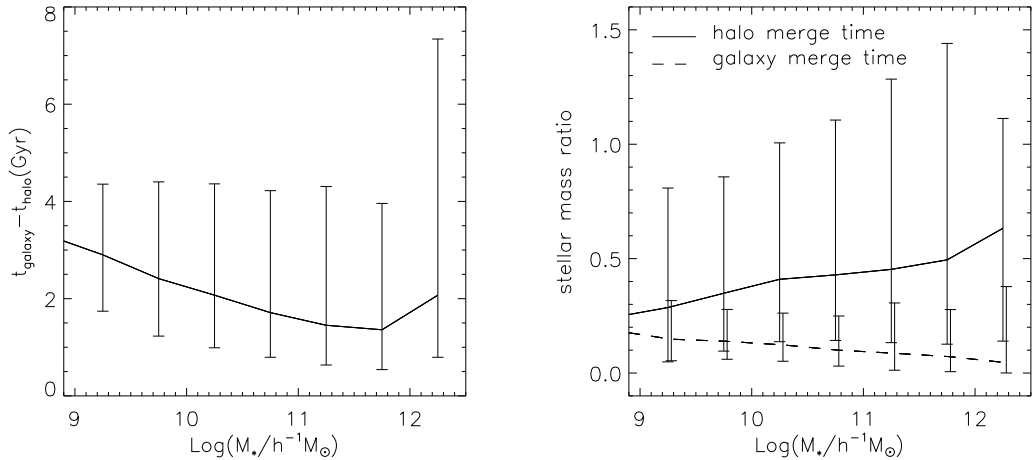}}\\%
\caption{ Some characteristics of the merging events with the
highest stellar mass ratios that take place during the history of a
galaxy: Left: the average time delay between the time that the
progenitor haloes merge and the time that the central galaxies
merge. Right: the stellar mass ratio of the galaxies at the time
that the  haloes merge (solid line) and at the time when the central
galaxies merge (dashed line). All results are plotted as a function
of the stellar mass of the central galaxy and error bars show the 68
percentile dispersion around the mean value. } \label{fig:massratio}
\ec
\end{figure*}

This is illustrated in Fig.~\ref{fig:massratio}. For every merging
event that occurs over the history of a galaxy, we record stellar
mass ratio information at the time when the progenitor haloes merge
and at the time when the galaxies themselves merge together. For
simplicity, we keep information for one randomly chosen merging
event in the history of each galaxy. In the left panel of
Fig.~\ref{fig:massratio}, we plot the average time that elapses
between the time when the two haloes merged and the time when the
galaxies themselves merged. Results are shown as a function of
galaxy stellar mass and the the error bars indicate 68 percentile
range in the distribution of delay times. As can be seen, the
typical delay time is around 2 Gyr, but individual time delays can
range between 1 and 5 Gyr. The delay times are typically shorter for
the progenitors of more massive galaxies.

In the right panel of Fig.~\ref{fig:massratio}, we plot the average
stellar mass ratios of the galaxies at the time when their haloes
merge (solid curve) and at the time when the two galaxies themselves
merge (dashed line). Notice that the stellar mass ratio can
sometimes be larger than 1; this happens when the galaxy inside the
smaller halo is more massive than the galaxy in the larger halo. As
we expect, the mass ratio of galaxies at the time when the galaxies
merge is smaller than that it is at the time when the haloes merge.
This effect is somewhat larger for the mergers that give rise to the
most massive galaxies at the present day.

\begin{figure}
\bc \hspace{-0.8cm}
\resizebox{9.cm}{!}{\includegraphics{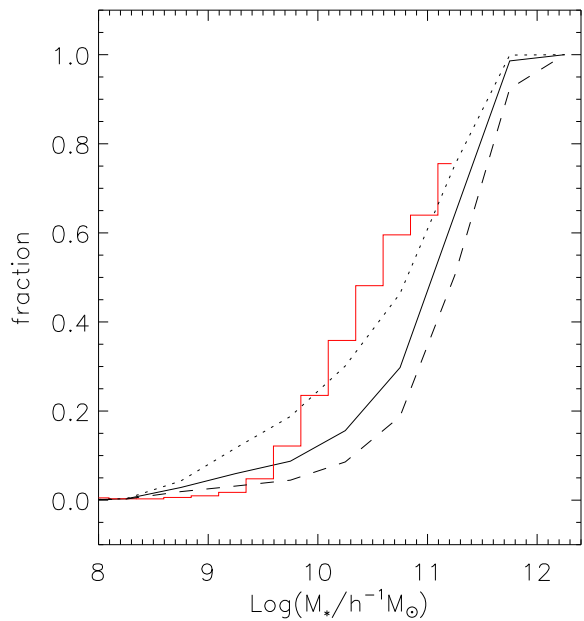}}\\%
\caption{ The solid curve shows the fraction of galaxies of given
stellar mass that are predicted to have experienced at least one
major merger. The red histogram shows the fraction of SDSS galaxies
with $z<0.06$ that are classified as AGN. The dotted and dashed
lines show the results from simulation when the mass ratio threshold
for defining major merger is changed to $0.2$ and $0.4$.}
\label{fig:sdss} \ec
\end{figure}

\subsection {Comparison with Observations}

In this section we have seen that the fraction of galaxies that have
experienced one or more major merging events is predicted to very
close to zero at stellar masses less than $\sim 10^{10} M_{\odot}$, but
a very steeply rising fraction of stellar mass for
$M_* > 10^{10} M_{\odot}$. We now compare this prediction with the
fraction of Sloan Digital Sky Survey  galaxies that contain an AGN.
We restrict the SDSS sample to redshifts $z< 0.06$ so that we are
still able to detect AGN with weak line emission (LINERs).
As shown by \citet{kauffmann2003}, weak-lined AGN become progressively
more difficult to identify at higher redshifts using SDSS spectra.
This is because these spectra are obtained through 3 arcsecond
diameter fibre apertures and the contribution from the stellar
population of the host galaxy becomes increasingly dominant in more
distant galaxies.

The results of the comparison are shown in Fig.~\ref{fig:sdss}. The
black curve shows the fraction of galaxies in the Millennium
Simulation of given stellar mass that have had at least one major
merger. The red histogram shows the fraction of galaxies in the SDSS
survey that are classified as AGN. As can be seen, both fractions
rise steeply from values close to zero at $M_* < 10^{10}
h^{-1}M_{\odot}$ to nearly unity at stellar masses greater than
$10^{11} h^{-1}M_{\odot}$. In Fig.~\ref{fig:sdss}, the dotted and
dashed lines show the results from simulation when the mass ratio
threshold for defining major merger is changed to $0.2$ and $0.4$.
Compared with the solid line where we use $0.3$ as the mass ratio
threshold to define a major merger, the increasing trends are about
the same for different thresholds in the range from $0.2$ to $0.4$.

\section{first black holes}
\label{sec:firstmerger}

\begin{figure*}
\bc
\hspace{-1.6cm}
\resizebox{16.cm}{!}{\includegraphics{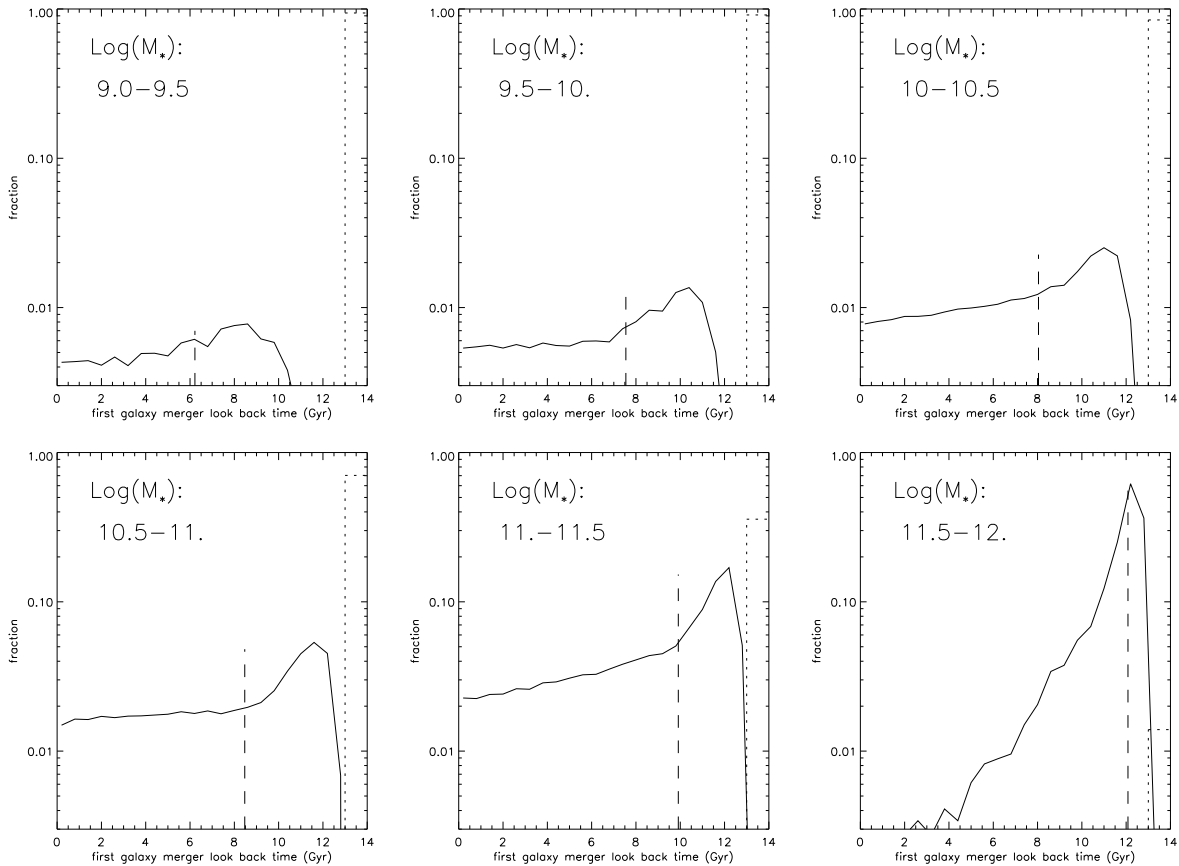}}\\%
\caption{
Distribution of time when a galaxy experiences its first major merger
for galaxies in different stellar mass bins (solid lines).
Galaxy stellar mass is plotted in units of
$h^{-1}M_{\odot}$. The vertical dashed lines show the median value
of the distributions. In each panel, the dotted bar shows the
fraction of galaxies that have never experienced a major merger.}
\label{fig:firstmerge}
\ec
\end{figure*}

In this section, we analyze when the first major merger that
produces the black hole in the galaxy is predicted to occur. In
Fig.~\ref{fig:firstmerge}, we plot the distribution of the times of
the first major merging events for galaxies with different
present-day stellar masses. The vertical dashed lines indicate the
median values of the distributions. The dotted bar in each panel
indicates the fraction of galaxies in each stellar mass bin that
have not experienced a major merger and are hence not included in
the distribution of merging times.

As can be seen, massive galaxies experience their first major
merging event at earlier epochs than less massive galaxies. Almost
no new black holes form in massive galaxies at the present day. The
distribution of black hole formation times in low mass galaxies is
much flatter. If the bulk of the black hole mass is built up in a
short period following the first major merger, this would explain
why present-day massive black holes have stopped growing, while low
mass black holes are still growing at a significant rate
\citep{heckman2004}.

\begin{figure}
\bc
\hspace{-0.8cm}
\resizebox{9.cm}{!}{\includegraphics{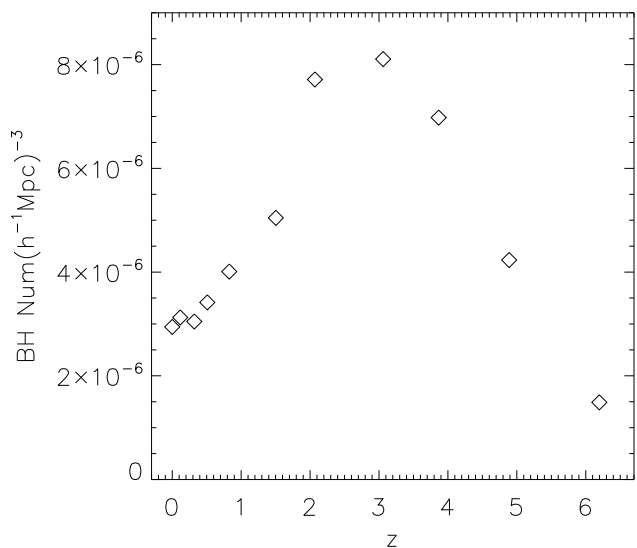}}\\%
\caption{ The number density of galaxies experiencing their first
major merger (diamonds) is plotted as a function of redshift. Each
merger is assumed to be visible for $10^7$ years. }
\label{fig:BHrate} \ec
\end{figure}

We now assume that the black holes formed from the first major
mergers of galaxies can shine and be observed for $10^7$ years. By
counting the numbers of such events at different redshifts, we can
compute the evolution in the number density of newly formed black
holes. The result is plotted as diamonds in Fig.~\ref{fig:BHrate}.
The comoving number density of such events peaks at redshift of z
$\sim2-3$, consistent with the observed peak in the number density
of bright quasars \citep{richards2006}. The decrease in the number
density of newly formed black holes to high redshifts is less
pronounced than that found by \citet{fan2001}, who show that the
luminous quasar density decreases by a factor of $\sim 6$ from
redshift $3.5$ to $5$. Note that we have not attempted to model the
predicted luminosity of the quasars in this work, so a direct
comparison with the observations is not possible. As we have
discussed, it is well possible that processes other than mergers
contribute to the low-luminosity quasar population.

\section{Conclusions}
\label{sec:discussions}

We analyze the merger histories of dark matter haloes and galaxies
in the {\em Millennium Simulation} and use our results to try
to understand the demographics of black holes in nearby galaxies.
Black holes are assumed to form only if a major merger occurs.
Although a significant fraction of low mass ($< 10^{10} M_{\odot}$)
galaxies have experienced minor mergers, less than a few percent are
predicted to have experienced a major merger.
If our assumption that a major merger is required in
order to form a black hole is correct, the majority of low mass
galaxies are predicted not to contain black holes
at the present day. This is one possible explanation of the observed
lack of AGN in low mass galaxies \citep{ho1997, kauffmann2003}.

We also investigate when galaxies of different stellar masses are predicted
to have formed their first black holes. High mass galaxies form their
first black holes at very early epochs. The distribution of formation
times is almost flat as a function of lookback time for
low mass galaxies. This means that if a low mass galaxy has a
black hole, there is a significant probability that it formed in the
last few Gigyears. We also compute the number density of newly formed
black holes as a function of redshift. We find that the peak number
density occurs at $z \sim2-3$, in good agreement with the observed peak
in the quasar space density. More detailed predictions for how AGN of
different luminosities are expected to evolve requires a more detailed
physical model for how the black holes accrete gas over the history of
the Universe. In addition, in certain wavebands AGN activity might be
obscured by gas and dust surrounding black hole \citep{hopkins2006}.
More detailed consideration of these issues will form the basis for
future work.

\section*{Acknowledgements}
We are grateful to Zuhui Fan, Gabriella De Lucia, Roderik Overzier
and Qi Guo for their detailed comments and suggestions on our paper.
Lan Wang would like to acknowledge the supports from NSFC under
grants 10373001, 10533010, and 10773001, and 973 Program (No.
2007CB815401). The simulation used in this paper was carried out as
part of the programme of the Virgo Consortium on the Regatta
supercomputer of the Computing Centre of the Max--Planck--Society in
Garching.

\bsp
\label{lastpage}

\bibliographystyle{mn2e}
\bibliography{BH}

\end{document}